\documentclass[article,reprint]{revtex4-1}

\usepackage{graphics}
\usepackage{graphicx}
\usepackage{amsmath}
\usepackage{amssymb}
\usepackage{bm}         

\renewcommand{\eqref}[1]{eq.~(\ref{eq:#1})}

\newcommand{\figref}[1]{Fig.~\ref{fig:#1}}
\newcommand{\Figref}[1]{Figure~\ref{fig:#1}}
\newcommand{\PUNT}[1]{}

\begin{document}

\title{Partially-disordered photonic-crystal thin films for enhanced
  and robust photovoltaics}

\author{Ardavan Oskooi$^\dagger$}
\email{oskooi@qoe.kuee.kyoto-u.ac.jp}

\author{Pedro A. Favuzzi$^\dagger$}

\author{Yoshinori Tanaka} 

\author{Hiroaki Shigeta}

\author{Yoichi Kawakami}

\author{Susumu Noda}

\affiliation{Department of Electronic Science \& Engineering, Kyoto
  University, Kyoto 615-8510, Japan}

\begin{abstract}
We present a general framework for the design of thin-film
photovoltaics based on a partially-disordered photonic crystal that
has both enhanced absorption for light trapping and reduced
sensitivity to the angle and polarization of incident radiation. The
absorption characteristics of different lattice structures are
investigated as an initial periodic structure is gradually
perturbed. We find that an optimal amount of disorder controllably
introduced into a multi-lattice photonic crystal causes the
characteristic narrow-band, resonant peaks to be broadened resulting
in a device with enhanced and robust performance ideal for typical
operating conditions of photovoltaic applications.
\end{abstract}


\maketitle

\noindent Recent interest in enhancing the light-trapping
functionality of thin-film photovoltaics has lead to a large number of
proposals for different nanostructured designs that exploit the wave
nature of incident
light~\cite{Bermel07,Zhou08,Chutinan09,Park09,Mallick10,Zhu10,Atwater10}. The
performance of these designs is typically optimized for only a limited
set of operating conditions involving a reduced bandwidth,
narrow-angular cone and single polarization. In this letter, we show
that by introducing a slight amount of disorder into a multi-lattice
photonic-crystal (PC) slab we can increase the light trapping over a
wideband spectrum while also reducing its angular and polarization
sensitivity. The presence of disorder itself may potentially make the
design more feasible for large-scale production. We demonstrate that
too much disorder, while improving the robustness of the design to
properties of the incident light, actually lowers the light trapping
relative to an optimum and we discuss the tradeoffs therein. The key
design principle, we argue, is to controllably introduce just the
right amount of disorder into the lattice so as to gradually broaden
the sharp resonances of the PC; retaining aspects of the high-peak
absorption while slightly reducing the dependence on specific-wave
features.

\begin{figure}
{\centering
  \resizebox*{1.0\columnwidth}{!}{\includegraphics{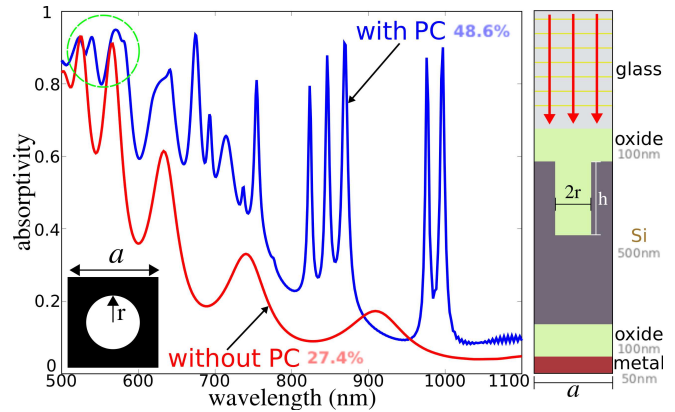}} \par}
\caption{Absorption versus wavelength profile at normal incidence for
  two thin-film photovoltaic designs: with a photonic crystal (PC,
  blue) and without (bare slab, red). The PC consists of a square
  lattice (periodicity, $a$=350nm) of circular rods ($r$=140nm,
  $h$=200nm) of oxide patterned in silicon (Si). The photon-absorption
  efficiency (from ~\eqref{efficiency}) is a measure of light trapping
  relative to a perfect absorber. The peaks circled in green are in
  the limit of strong Si absorption and thus depend mostly on film
  thickness, not lattice design.}
\label{fig:compare_designs}
\end{figure}

The delicate interference effects that underlie wave-optics designs
for photovoltaics often gives rise to performance that is sensitive to
the wave properties and thus makes it difficult to obtain enhanced
light-trapping over a wideband spectrum spanning multiple angles and
polarizations. A variety of different PC designs have already been
proposed for improved light-trapping in thin-film
photovoltaics~\cite{Zhou08,Chutinan09,Park09,Mallick10}. The core
principle of such designs are a collection of narrow-band
resonances~\cite{Imada99,Notomi00,Noda01} based on the coherent
scattering of photons~\cite{Joannopoulos08} that together contribute
to broadband absorption enhancement relative to an unpatterned slab of
a given absorbing media (usually silicon or gallium
arsenide). ~\Figref{compare_designs} shows the absorption profile for
two thin-film photovoltaic designs consisting of a 500nm-thick silicon
(Si) absorbing layer, with and without a PC, having a metal
back-reflector. The PC is formed by patterning a square lattice
(periodicity 350nm) of oxide (refractive index 1.8) rods (radius
140nm, height 200nm) within the Si film (a cross section of the unit
cell is shown in the schematic of ~\figref{compare_designs}). These
specific design parameters for the PC, while not optimized for photon
absorption over the given spectrum (quantified in the next section),
represent a valid starting point for exploring the extent to which
variations to the unperturbed lattice can improve light absorption. No
front- or rear-surface texturing~\cite{Ghebrebrhan09,Sheng11} is
employed to boost scattering into the guided modes of the slab as we
are only interested in the effect of the nanostructured lattice on the
light trapping. A normally-incident (later, we will consider a range
of oblique angles) planewave spanning the wideband spectrum from 500nm
(Si is strongly absorbing below 500nm which renders less important the
effect of the lattice to boost light trapping) to 1100nm (the indirect
bandgap of Si corresponds to a wavelength of 1107nm) impinges on the
structure from above through a glass ($n$=1.5) substrate. We use an
open-source finite-difference time-domain (FDTD) simulation
tool~\cite{Oskooi10} to compute the fractional absorptivity of the Si
film. The absorption spectra for the PC shows narrow, high-amplitude
peaks - signatures of the coherent--resonant-Bloch modes - whereas the
unpatterned slab has broad Fabry-P\'{e}rot resonances with low
amplitude.

We can quantify the light-trapping efficiency of each design relative
to an ideal perfect absorber (unity absorptivity over the wavelength
interval of interest) by assuming that each absorbed photon with
energy larger than the bandgap of crystalline Si generates an exciton
which contributes directly to the short-circuit
current~\cite{Ghebrebrhan09}. This corresponds to the following
definition of photon-absorption efficiency:

\begin{equation}
\frac{\int_{500nm}^{1100nm}\lambda\frac{dI}{d\lambda}A(\lambda)d\lambda}{\int_{500nm}^{1100nm}\lambda\frac{dI}{d\lambda}d\lambda},
\label{eq:efficiency}
\end{equation}

where $dI/d\lambda$ is the terrestrial power density per unit
wavelength from a blackbody at a temperature of 5778K~\cite{ASTM05}
and $A(\lambda)$ is the absorptivity computed from the
simulations. ~\Figref{compare_designs} shows that the introduction of
a PC lattice into the absorbing Si layer results in a near doubling of
the overall efficiency compared to an unpatterned slab (48.6\%
vs. 27.4\%). Somewhat non-intuitively, the effect of \emph{removing}
the absorbing material (Si) to form the PC lattice has been to
\emph{enhance} the absorption via an increase in light
localization. The sharp peaks of the PC (corresponding to a large Q
factor~\cite{Joannopoulos08}) indicate that coupling into (and out of)
the modes is challenging and will depend strongly on the incident-wave
conditions which we will verify later on. While a patterned design
seems to be better for light trapping than an unpatterned one, what is
less clear is whether the individual, narrow-band, large-amplitude
peaks can be somehow broadened to further boost the overall absorption
and the extent to which a given lattice design can be made robust with
respect to different angles and polarizations of incident light.

\begin{figure}
{\centering
  \resizebox*{1.0\columnwidth}{!}{\includegraphics{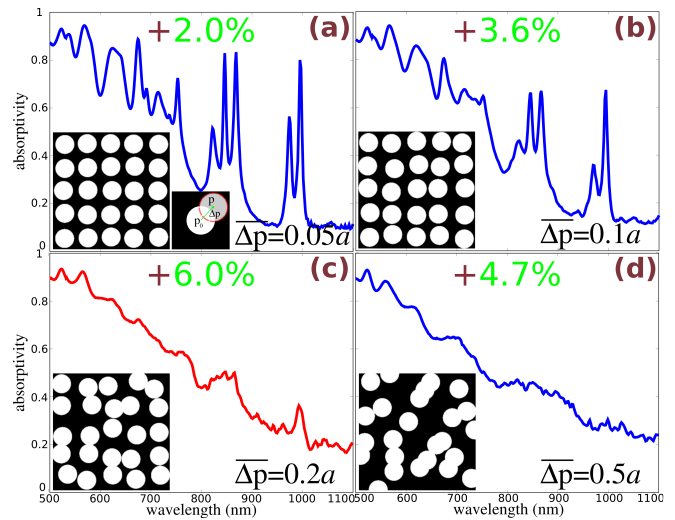}} \par}
\caption{The evolution of the absorption profile for a
  gradually-increasing amount of positional disorder introduced into
  the PC. Each rod is perturbed from its position in the unperturbed
  10x10 lattice by an amount $\Delta p$ chosen randomly from a uniform
  distribution of values between 0 and $\overline{\Delta p}$. The
  efficiency values, in green, are the absolute change relative to the
  unperturbed lattice. Insets a, b and c (in red, with maximal light
  trapping) show traces of the Bloch-mode resonances while inset d is
  Anderson localized with sub-optimal absorption.}
\label{fig:vary_pos_disorder}
\end{figure}

The complex nature of the wave scattering that gives rise to the
Bloch-mode resonances of the PC lattice depends delicately on
perfect-structural order. It is natural to wonder then how the
absorption spectra changes as a gradual amount of disorder is
introduced into the lattice. ~\Figref{vary_pos_disorder} shows the
evolution of the absorption profile for a steadily increasing amount
of positional disorder. Each rod is perturbed from its position in the
unperturbed lattice, a supercell consisting of 10x10 unit cells, by an
amount $\Delta p$ chosen randomly from a uniform distribution of
values between 0 and $\overline{\Delta p}$ for both orthogonal
in-plane directions. Two separate simulations are made for each
structure and the results averaged due to the random nature of the
design. As the disorder progressively increases ($\overline{\Delta
  p}$=0.05$a$, 0.1$a$, 0.2$a$, 0.5$a$), the peaks broaden more and
more, converging to a profile resembling the unpatterned slab
(~\figref{compare_designs}, in red) but with higher absorption at each
wavelength (peak broadening due to fabrication-induced disorder was
first reported in a study by Chutinan and John~\cite{Chutinan08} but
which did not investigate the effect of a controlled amount of
disorder built into the design itself). In the limit of large
disorder, no trace of the coherent scattering remains as the
transition has been made to Anderson
localization~\cite{Schwartz07,Toninelli08}. Note that a
\emph{transformation} in the qualitative behavior of the absorption
spectra occurs as the lattice changes from $\overline{\Delta
  p}$=0.2$a$ (inset c) to $\overline{\Delta p}$=0.5$a$ (inset d): in
the former, traces of the resonant modes are still visible while in
the latter they have completely disappeared. This suggests that, upon
the addition of a relatively small amount of disorder, the wave
behavior is still governed more or less by the scattering of
Bloch-mode resonances while too much disorder results in a very
different behavior marked by the incoherence of Anderson
localization~\cite{Rechtsman11}. ~\Figref{vary_pos_disorder} also
shows that a modest amount of disorder ($\overline{\Delta p}=0.2a$)
maximizes the overall efficiency of the design (a 6.0\%
\emph{absolute} increase beyond the unperturbed lattice) but that too
much disorder ($\overline{\Delta p}>0.5a$) actually lowers the
efficiency gains. A similar observation was previously made in the
study of an inverse-opal structure~\cite{Toninelli08} but the notion
of a partially-disordered PC has not yet been extended to
light-trapping in thin films for photovoltaic applications. The effect
of disorder on light trapping is twofold: 1) it can be used to tune
the density of states (DOS), specifically the ratio of Bloch to
Anderson-localized modes and 2) it can be used to control the coupling
rate into and out of the lattice. And hence, the result of little
disorder is likely to increase the DOS slightly~\cite{Chutinan08} and
also improve the coupling efficiency leading to an overall boost in
the light trapping whereas large disorder eliminates the coherent
modes entirely and enables the light to leak out as readily as it can
couple in thus lowering the performance. A careful understanding of
the role that disorder plays in these complimentary aspects of DOS and
mode coupling may potentially lead to designs with better
light-trapping behavior.

\begin{figure}
{\centering
  \resizebox*{0.9\columnwidth}{!}{\includegraphics{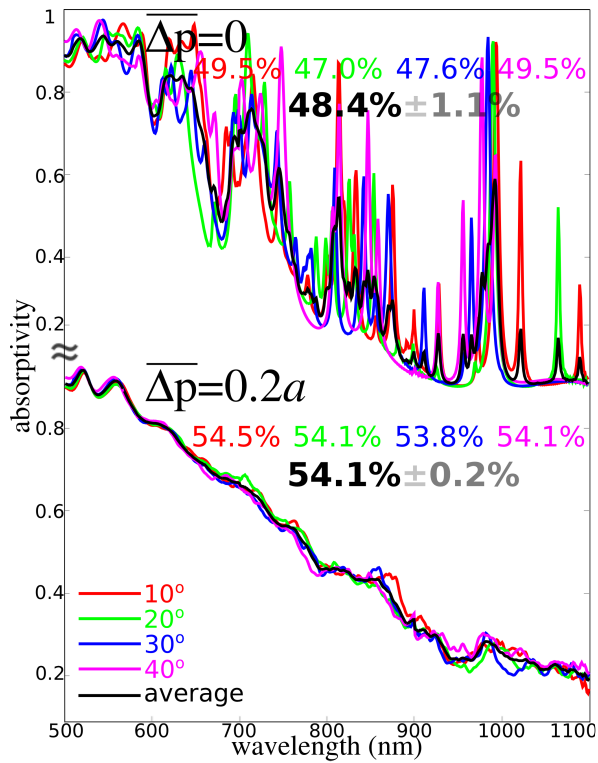}} \par}
\caption{Absorption profile for two different structures (the original
  unperturbed PC and {one with optimal} disorder) at four off-normal
  angles of incidence with the TE polarization. The absorption profile
  for the individual angles are colored while the averaged profile is
  shown in black. Note large increase in robustness with disorder
  (i.e. standard deviation of the averaged efficiency decreases
  sharply). The TM polarization data (not shown) has similar
  characteristics.}
\label{fig:vary_angle}
\end{figure}

In addition to the large efficiency improvement of a
partially-disordered lattice compared to an unperturbed one, the
slight broadening of the peaks offers another significant advantage:
it reduces the dependence of the localized modes on the incident-wave
properties. Up to this point, the absorption spectra have only been
shown for the case of a normally-incident planewave source. At
off-normal incidence, the polarization degeneracy is lifted enabling
us to better investigate the dependence of a given design's efficiency
on the wave-source conditions. ~\Figref{vary_angle} shows the
absorption profile for two structures - an unperturbed PC
($\overline{\Delta p}$=0) and one with optimal disorder
($\overline{\Delta p}$=0.2$a$) - at four different angles (10$^\circ$,
20$^\circ$, 30$^\circ$, 40$^\circ$) for the TE polarization (electric
field in the plane of incidence). Once again, we perform two
simulations for the random structure and average the results to
increase accuracy. As shown in the figure, the resonant peaks of the
unperturbed PC have a strong dependence on the angle of the incident
planewave. Moreover there is a corresponding variation in the
absorption efficiency from 49.5\% to 47.0\% which underscores just how
sensitive the PC design's performance is to external factors (a
lattice optimized for absorption at normal incidence showed an even
larger range of variation from 53.1\% to 47.3\%). The
partially-disordered optimal structure, on the other hand, shows much
less sensitivity (quantified by a standard deviation value of the
averaged efficiency that is more than a factor of five smaller) while
a more-disordered structure has even less sensitivity; yet it is
(expectedly) at the expense of a slightly-lower overall efficiency
(the trend is also true for the TM polarization). These results
suggest that the choice of disorder strength requires a careful
consideration of the trade off between enhancing the light-trapping on
the one hand and making the structure more robust to angle and
polarization on the other: too little disorder and the design
performance depends sensitively on outside conditions, while too much
disorder lowers the efficiency relative to an optimal amount (albeit
any disordered lattice is superior to the unpatterned slab). Such an
insight into the light-localization phenomenon of disordered
structures points the way towards an interesting design strategy: find
the \emph{maximum} amount of disorder that still preserves some
fundamental Bloch character of the modes and retains features of the
large-peak amplitudes to improve light trapping and robustness without
broadening too much so as to be completely incoherent and leaky.

\begin{figure}
{\centering
  \resizebox*{0.9\columnwidth}{!}{\includegraphics{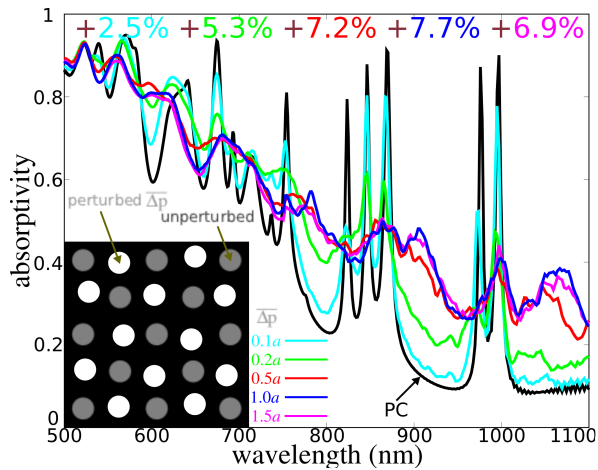}} \par}
\caption{Absorption profile, at normal incidence, for a multi-lattice
  structure composed of two independent sublattices (white and grey)
  where only one is perturbed ($\overline{\Delta p}$). Data for five
  different structures are shown (in color) with the unperturbed PC
  (in black) added for reference. Efficiency values for the five
  perturbed multi-lattice structures are the absolute change relative
  to the unperturbed lattice.}
\label{fig:correlated_random}
\end{figure}

An important design consideration then, stemming from the observations
made above, is to find a structure where the disorder (of different
types and length scales) has both a small footprint and can be
precisely controlled in order to better combine aspects of the
high-absorption peaks of the unperturbed PC with the broadened,
polarization- and angle-independent peaks of the strongly-disordered
lattice. To investigate this point in more detail, we now examine a
multi-lattice design that consists of \emph{two} sublattices where the
positional disorder of each sublattice can be independently tuned (the
radius of all rods are equal). This multi-lattice structure, shown in
~\figref{correlated_random}, is an example of a lattice design that
should offer a finer degree of control for introducing an optimal
amount of disorder into the unperturbed structure to obtain the
desired characteristics especially when compared to the single-lattice
structure where all rods are perturbed in the same
way. ~\Figref{correlated_random} shows the absorption spectra, at
normal incidence, of five different multi-lattice structures with
steadily increasing disorder where \emph{only} one of the two
independent sublattices is perturbed. Indeed the broadening of the
peaks happens gradually enough, acting to increase the absorptivity in
regions where it was nearly negligible in the unperturbed lattice
while simultaneously better preserving traces of the high-peak
resonances, that the light-trapping efficiency can be made to
\emph{surpass} that of the optimal single-lattice structure of
~\figref{vary_pos_disorder} (in this example, the multi-lattice
structure shows an efficiency gain of 7.7\% at a $\overline{\Delta p}$
of 1.0$a$). (Note that since the disorder footprint is now smaller
than before that more disorder is required to achieve similar gains
and beyond.) By merging two sublattices into one structure this way,
we are able to more effectively blend the large--peak-amplitude
features of the coherent Bloch scattering with the broadening due to
disorder.

While we demonstrated peak-broadening behavior using only positional
disorder, the same phenomenon was observed for radial disorder
although for a fixed supercell size the latter disorder type does not
preserve the quantity of Si thus somewhat complicating the
analysis. We also investigated the effect of disorder starting with an
\emph{optimized} unperturbed PC lattice (53.2\% at $a$=500nm and
$r$=150nm) and discovered similar trends of increasing and then
decreasing absorption gains, although the maximum efficiency
improvement was small by comparison (only 2.2\% at a $\overline{\Delta
  p}$ of 0.05$a$). This might suggest that ultimately the optimal
amount of disorder, different for each unperturbed starting lattice,
will result in nearly--identical maximum efficiencies (in this case
around 55\%, likely limited by the material properties and film
thickness of Si) and that the main advantage of using a non-optimal
starting lattice is simply that more disorder can be incorporated
therefore making the device more robust. Minor fluctuations in the
absorption spectra, particularly in the long-wavelength regime for
large disorder, are strictly numerical noise due to the
finite-supercell size and periodic-boundary conditions. In fact, as
the supercell size is increased (and with it the effect of
randomness), these small oscillations diminish and the absorption
profile becomes smoother. Another approach, more relevant for reducing
noise in smaller-wavelength features, would be to average the results
over a larger number of random iterations (currently two). In addition
to a uniform distribution for our random model, we also investigated a
normal distribution where the mean positional disorder is fixed at 0
and the standard deviation is varied. The only significant difference,
indeed a major one, in the absorption spectra between the two
distribution models was that the uniform distribution lead to slightly
less broadening due to their being an upper limit for the
perturbation. It therefore appears that a uniform model of randomness
is better at controlling the disorder than a normal one. A more
comprehensive investigation of the role of different lattice designs
and random-distribution models will be undertaken in a future
publication.

In summary, we have described a design principle for enhancing both
light trapping and robustness using a partially-disordered
photonic-crystal multi-lattice structure where the amount of disorder,
a parameter controlling the DOS as well as the coupling rate between
external and internal modes, has both a small footprint and can be
carefully chosen. As the performance of typical wave--optics based
designs for photovoltaic applications depends sensitively on
structural order and external conditions, this strategy may provide a
route towards the realization of thin-film photovoltaics with high
performance over a wide range of operating criteria.

This work was supported by Core Research for Evolutional Science and
Technology (CREST) from the Japan Science and Technology Agency and
the Japan Society for the Promotion of Science (JSPS).

\end{document}